\documentclass[letterpaper, 10pt,conference]{ieeeconf}

\usepackage{graphicx}          

\usepackage{amsmath,amssymb,amsfonts}
\usepackage{graphicx}
\usepackage{textcomp}
\usepackage{algpseudocode}
\usepackage{bm}
\usepackage{color}

\newtheorem{remark}{Remark}

\newcommand{\rr}{{\mathbb R}}

\renewcommand{\thesection}{\arabic{section}}
\renewcommand{\theequation}{\thesection.\arabic{equation}}
\newcommand{\beq}[1]{\begin{equation} \label{#1}}
\newcommand{\eeq}{\end{equation}}
\newcommand{\bea}{\bed\begin{array}{rl}}
\newcommand{\eea}{\end{array}\eed}
\newcommand{\bed}{\begin{displaymath}}
\newcommand{\eed}{\end{displaymath}}
\newcommand{\barray}{\begin{array}{ll}}
\newcommand{\earray}{\end{array}}

\newcommand{\disp}{\displaystyle}

\newcommand{\al}{\alpha}

\newcommand{\wdt}{\widetilde}

\def\para#1{\vskip 0.4\baselineskip\noindent{\bf #1}}

\def\rr{{\Bbb R}}
\def\({\left(}
\def\){\right)}
\def\one{{\hbox{1{\kern -0.35em}1}}}

\IEEEoverridecommandlockouts
\begin{document}

\thispagestyle{empty}
\pagestyle{empty}

\title{\LARGE \bf
Deep Filtering with DNN, CNN and RNN}


\author{ Bin Xie$^{1}$, Qing Zhang$^{2}$
\thanks{$^1$ Mathematics Department, University of Georgia, Athens, GA 30602 {\tt\small bin.xie@uga.edu} }
\thanks{$^2$ Mathematics Department, University of Georgia, Athens, GA 30602 {\tt\small qz@uga.edu}}
}

\maketitle

\begin{abstract}                          
   This paper is about a deep learning approach for general
   filtering. The idea is to train a neural network with Monte Carlo
   samples generated from a nominal dynamic model.
   Then the network weights are applied to Monte Carlo samples
   from an actual dynamic model.
   A main focus of this paper is on the deep filters with three major neural
   network architectures (DNN, CNN, RNN). 
Our deep filter compares
  favorably to the traditional Kalman filter in linear cases and outperform
  the extended Kalman filter in nonlinear cases.
  Then a switching model with jumps is studied to show the adaptiveness
  and power of our deep filtering.
  Among the three major NNs, the CNN outperform the others on average.
  while the RNN does not seem to be suitable for the filtering problem.
    One advantage of the deep filter is its robustness
  when the nominal model and actual model differ.
  The other advantage of deep filtering is real data can be used directly to
  train the deep neutral network. Therefore, model calibration can be by-passed
  all together.
 
\end{abstract}


\section{Introduction}

This paper is about a deep learning method for filtering with
different neural networks.
Filtering is concerned with state estimation in systems
that are not completely observable. There are many applications
of filtering in engineering and financial systems.
Traditional approaches focus on least squares
estimators with Gaussian distributions.
The corresponding filtering problem is to find the
conditional mean of the state given the observation up to time $n$.
The best known filter is the Kalman filter for linear models.
For some nonlinear models,  extended Kalman filters provide good
approximations.
We refer the reader to Anderson and Moore \cite{AndersonM} for
background and technical details.

Classical results in nonlinear filtering can be found in
Duncan \cite{Duncan} which focuses on conditional densities for diffusion
processes; Mortensen \cite{Mortensen} that deals with the
most probable trajectories;
Kushner \cite{Kushner}, which derives nonlinear filtering equations,
and Zakai \cite{Zakai}, which treats equations of unnormalized
conditional expectations.

Following these classical work,
there are much progresses made in the past decades.
For instance, hybrid filtering can be found in
Hijab \cite{Hijab-paper} with an unknown constant,
Zhang \cite{Zhang-switch} with small observation noise,
Miller and Runggaldier \cite{MillerR} with Markovian
jump times,
Blom and Bar-Shalom \cite{BlomB} for discrete-time hybrid model
and the Interactive Multiple Model algorithm,
Dufour et al \cite{DufourBE1,DufourBE2} and
Dufour and Elliott \cite{DufourE} for models with regime switching.
Some later developments along this line can be found in
Zhang \cite{Zhang-switch,Zhang1,Zhang2}.
Despite these important progresses, the computation of filtering remains a
daunting task. For nonlinear filtering, there have been yet
feasible and efficient  schemes to mitigate
high computational complexity (with infinite dimensionality). Much effort
has been devoted to finding computable approximation schemes.

Recently, Wang et al. \cite{WangYZ} developed a deep learning framework for
general filtering.
Given a partially observed system, the idea is to
train a deep neutral network using Monte Carlo samples generated from
the system.
The observation process is taken as inputs to the dense neural network and the 
state from the Monte Carlo samples is used as the target.
A least squares loss function of the target and calculated output
is to be minimized to determine the weight vectors. Then these weight vectors
are applied to a separate set of Monte Carlo samples of the
actual dynamic model. Such a state estimation procedure is termed deep filter.
In \cite{WangYZ}, computational experiments with linear, nonlinear,
switching models have shown the adaptiveness, robustness, and effectiveness of
the deep filter.
  A major advantage of the DF is its robustness
  when the nominal model and actual model differ.
  Another advantage of the deep filtering is that
  real-world data can be used directly to train
  the deep neutral network. Therefore, model calibration is no longer needed
  in applications.

 We point out that the experiments reported in  Wang et al. \cite{WangYZ} were only 
 conducted with DNNs. It would be interesting to examine how the
 filtering schemes work under different neural networks.
 It is the purpose of this paper to further our studies of deep filtering
 under convolution neural network (CNN) and recurrent neural network (RNN).
 We study both linear and non-linear systems and
 compare the performance of the DF under these different neural networks.
 We show that the deep filter compares
  favorably to the Kalman filter in linear cases and outperform
  the extended Kalman filter in nonlinear cases.
  A switching model with jumps is studied to show the adaptiveness
  of the deep filtering.
  Among the three major NNs, the CNN outperform the others on average.
  While the RNN does not seem to be suitable for the filtering problem.
    
 The rest of this paper is arranged as follows.
 In Section 2, we present our methodology and various neural networks
 under consideration. In Section 3, we report our computational experiments.
 Then in Section 4, we conclude the paper with remarks.




\section{Methodology}

Let $x_n \in \rr^{m_1}$ denote the state process with system equation
\beq{sys-eqn}
x_{n+1}=f_n(x_n,u_n),\ x_0=x,\ n=0,1,2,\ldots,
\eeq
for some suitable functions $f_n: \rr^{m_1} \times \rr^{l_1} \mapsto \rr^{m_1} $ and system noise $\{u_n\}$ with $u_n \in \rr^{l_1}$.
A function of $x_n$ can be observed with possible noise corruption.
In particular, the observation process $y_n\in \rr^{m_2}$ is given by
\beq{obs-eqn}
y_n=h_n(x_n,v_n),
\eeq
with noise $\{v_n\}$, $v_n \in \rr^{l_2}$, and $h_n: \rr^{m_1} \times \rr^{l_2} \mapsto \rr^{m_2}$.

\para{Deep Filter.}
The DF for (\ref{sys-eqn}) and (\ref{obs-eqn}) can be given as follows:
Let $N_{\rm seed}$ denote the number of training sample paths and
let $n_0$ denote the training window for each sample path.
For any fixed $\kappa=n_0,\ldots,N$
with a fixed
$\omega$, we take
$\{y_\kappa(\omega),y_{\kappa-1}(\omega),\ldots,y_{\kappa-n_0+1}(\omega)\}$
as the input vector to the neural network and $x_\kappa(\omega)$ as the target. In what follows, we shall suppress the $\omega$ dependence.
Fix $x_\kappa$,
let
$\xi_\ell$
denote the neural network output at iteration $\ell$, which depends on the parameter $\theta$.
Our goal is to
find an NN weight $\theta \in \rr^{m_3}$ so as to minimize
the loss function
\beq{loss}
L(\theta)={\frac{1}{2}} E
 |\xi_\ell- x_\kappa|^2.
\eeq

We use the
stochastic gradient descent (SGD) method to find the NN parameters $\theta$ by minimizing the loss function with learning rate $\gamma \in (0,1) $.
\beq{sgd}
\theta _{n+1} = \theta _n - \gamma \displaystyle \frac{\partial L (\theta_n)}{\partial \theta}.
\eeq

Then these weights $\theta$ are used to out-of-sample
data
$\{\check\omega\}$ with
the actual observation $y_n(\check\omega)$ as inputs in
the subsequent testing stage which leads to neural network output $\wdt x_n(\check\omega)$.
Such a state estimation procedure is called deep filter.

\subsection{Network Specifications}


Typical neural networks for deep learning applications include
dense neural networks (DNN), convolutional neural networks (CNN),
and recurrent neural networks (RNN).

\begin{figure}[thpb]
	\begin{center}
		\includegraphics[width=0.8\linewidth]{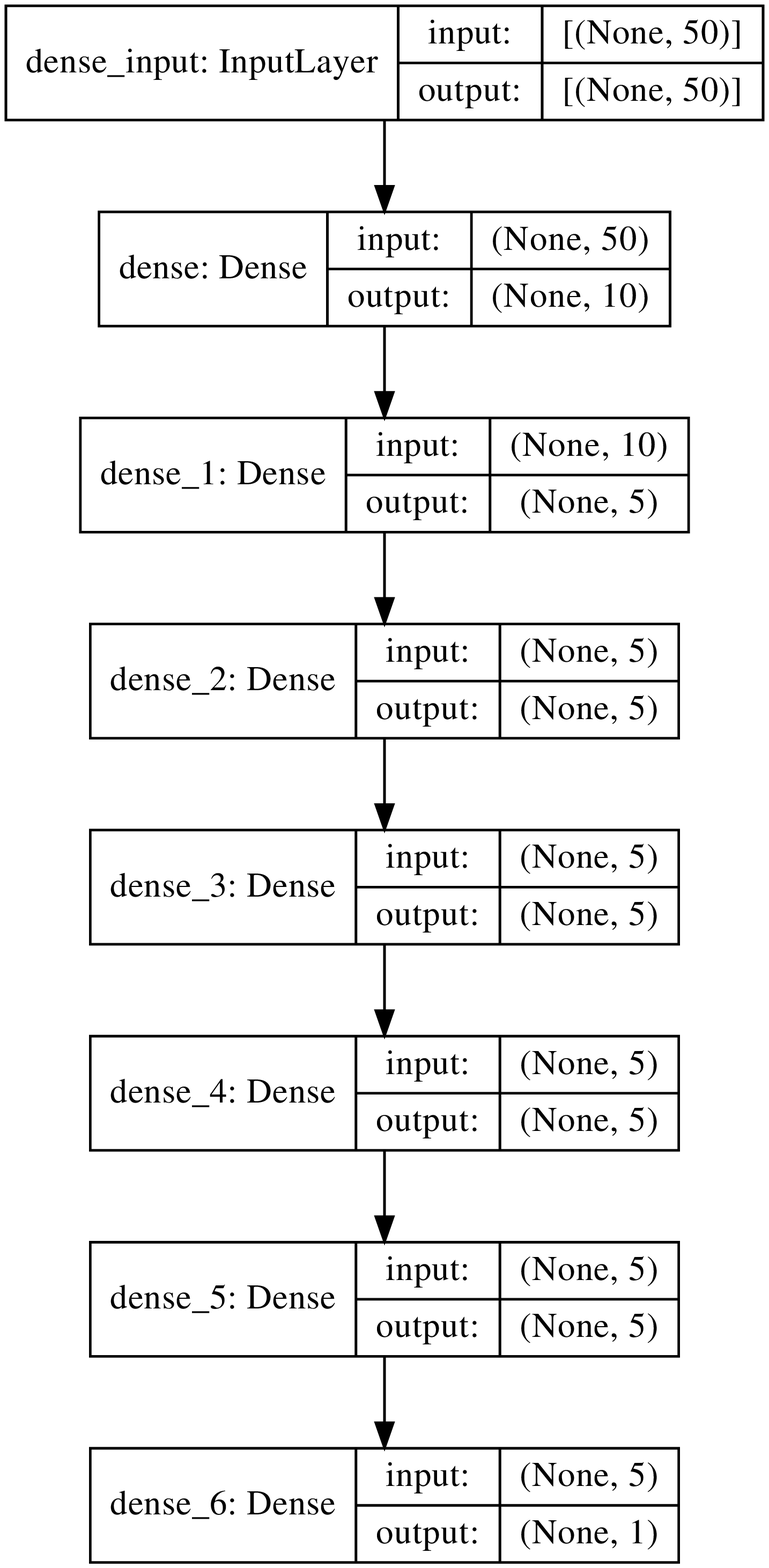}
		\caption{The DNN Structure }%
		\label{fig:2ddnn}
	\end{center}
\end{figure}

In this paper, we first consider the fully connected DNNs.
As in  Wang et al. \cite{WangYZ}, we use the DNN with 5 hidden layers (each with 5 neurons).
We also use the sigmoid activation function for 
all hidden layers and the simple activation for the output layer.
Figure \ref{fig:2ddnn} shows the DNN structure used in this paper.

\begin{figure}[thpb]
	\begin{center}%
		\includegraphics[width=0.8\linewidth]{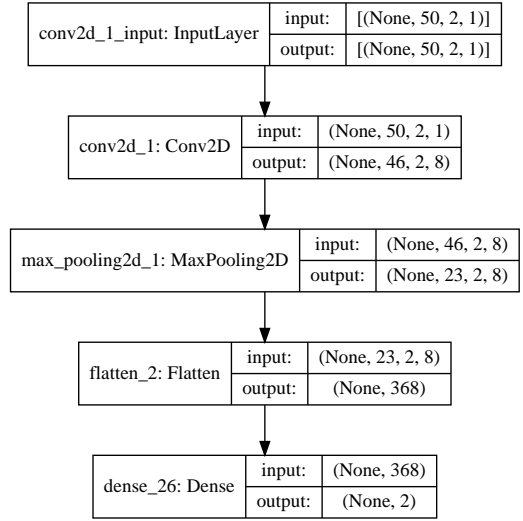}
		\caption{The CNN Architecture}
		\label{fig:cnn2d}
	\end{center}
\end{figure}

A CNN is another architecture of NN and it allows to model both time and
space correlations in multivariate signals. 
A CNN utilizes convolution operation to seize the local features
with several kernel filters.
CNNs are widely used in image classification, image recognition, and
computer vision. 
Figure \ref{fig:cnn2d} shows the CNN structure used in this paper.

\begin{figure}[thpb]
	\begin{center}
		\includegraphics[width=0.8\linewidth]{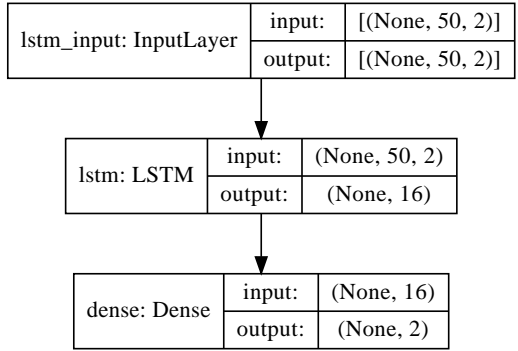}
		\caption{The LSTM Architecture}
		\label{fig:lstm}
	\end{center}
\end{figure}

An RNN is the other architecture of neural network where connections between
nodes form a directed graph along a temporal sequence.
A CNN is of feed-forward NN that uses filters and pooling layers,
whereas a RNN feeds results back into the network.
A RNN is often used in speech recognition.
A popular RNN architecture is the long short-term memory (LSTM).
It can process not only single data points, but also entire sequences of data.
By setting logic gates in a calculation cell, the LSTM network collect
efficient information through the time series.
Figure \ref{fig:lstm} shows the RNN structure used in this paper. 
For further details on related neural networks, we refer the reader to
Chollet \cite{chollet2018deep}.


\section{Numerical Experiments}

Following the system equations (\ref{sys-eqn}) and (\ref{obs-eqn})
for $(x_n,y_n)$, we generate 
$N_{\rm seed}=5000$ Monte Carlo sample paths (in-sample)
for the neural network training. Then we generate a separate set of
$N_{\rm seed}=5000$ samples (out-of-sample) to test the results.
We take the time horizon $N=1000$ and use the window size  
$n_0=50$ of input to train the network.
We use the stochastic gradient decent algorithm with learning rate $\gamma=0.1$.

The following relative error is used in this paper.
Given vectors
$\xi^1(\omega)=(\xi^1_{n_0}(\omega),\ldots,\xi^1_N(\omega))$
and $\xi^2(\omega)=(\xi^2_{n_0}(\omega),\ldots,\xi^2_N(\omega))$, define

\[
\Vert \xi^1-\xi^2\Vert=
\frac{\disp \sum_{n=n_0}^{N}\sum_{m=1}^{N_{\rm seed}}
  |\xi^1_n(\omega_m)- \xi^2_n(\omega_m)|}{N_{\rm seed}(N-n_0+1){\Sigma}},
\]
where
\[
{\Sigma}=\frac{\disp\sum_{n=n_0}^{N}\sum_{m=1}^{N_{\rm seed}}
(|\xi^1_n(\omega_m)|+|\xi^2_n(\omega_m)|)}{N_{\rm seed}(N-n_0+1)}.
\]

In this paper, we work with linear, non-linear, and switching models to
test deep filtering schemes. We start with linear systems.

\para{Linear Systems.}
We consider the linear system
\beq{linear-sys}
\left\{\begin{array}{l}
x_{n+1}=F_n x_n+G_n u_n,\ x_0=x,\\
y_n=H_n' x_n+v_n, \ n=0,1,2,\ldots,
\end{array}\right.
\eeq

for some matrices $F_n$, $G_n$, and $H_n$ of appropriate dimensions.
Here $u_n$ and $v_n$ are independent random vectors having
Gaussian distributions with mean zero and
$E(u_nu_l')=\delta_{nl}I$, $E(v_nv_l')=\delta_{nl}I$,
for $n,l=0,1,2,\ldots$, where $\delta_{nl}=1$ if $n=l$ and $0$ otherwise.

\para{1D Linear Model.}
In particular, the one-dimensional system is given as follows :
\beq{1dlinequ}
\left\{\begin{array}{l}
x_{n+1}=(1+0.1\eta)x_n+\sqrt{\eta}\ \sigma u_n,\ x_0=1,\\
y_n= x_n+\sigma_0 v_n, \ n=0,1,2,\ldots,
\end{array}\right.
\eeq
with $u_n$ and $v_n$ which are independent Gaussian $N(0,1)$ random variables.
We take $\sigma=0.7$, $\sigma_0=0.5$, and the step size $\eta=0.005$.

\para{2D Linear Model.}
In two-dimensional setting, we consider the system
\beq{2dlinequ}
\left\{\begin{array}{l}
x_{n+1}=(1+0.1\eta F^0_n)x_n+\sqrt{\eta}\  G^0_n u_n,\ x_0=[1,1],\\
y_n=H'_n x_n+\sigma_0 v_n, \ n=0,1,2,\ldots,
\end{array}\right.
\eeq
with $u_n$ and $v_n$ being independent 2D normal random variables.

We take $\eta=0.005$, $\sigma_0=0.5$, 
$G^0_n=I$, $H_n=I$ and
$F^0_n=\big(\begin{smallmatrix}0 & 1\\1 & 1 \end{smallmatrix}\big)$.

With these parameters, the corresponding numerical relative errors are
included in Table~\ref{tab:extlin}. The DF with neural networks
perform comparable to the KF except the one
with the RNN in two-dimensional case.
The DF with the CNN produces a little better relative errors.


\begin{table}[thpb]
\centering
\begin{tabular}{|l|l|l|l|l|}
\hline
 & KF  & DNN  & CNN  & RNN \\ \hline
1D Linear  &4.17  & 4.51 & 4.17 &  4.50 \\ \hline
2D Linear  &4.45  & 4.43 & 4.24 &  8.26    \\ \hline
\end{tabular}
\caption{Relative Errors (in \%) for Linear Models}
\label{tab:extlin}
\end{table}

\para{Robustness of Deep Filtering.}
In this section, we examine the robustness of deep filtering.
We consider separately the nominal model and the actual model.
A nominal model (NM) is an estimated model.
It deviates from real data for different applications.
In this paper, it is used to train our NNs, i.e.,
a selected mathematical model is used to
generate Monte Carlo sample paths to train the NN.
The coefficients of the mathematical model are also used
in Kalman filtering equations for comparison.

An actual model (AM), on the other hand, is about
the simulated (Monte Carlo based)
environment.
It is used in this paper for testing purposes.
In real world applications, the observation process is the
actual process obtained from real physical process.
To test the model robustness, we consider the case when the NM's observation
noise differs from the AM's observation noise.


To examine the robustness of the DF, we fix $\sigma_0=\sigma_0^{NM}=2$ and
vary $\sigma_0=\sigma_0^{AM}$. The relative errors for the
one-dimensional model case are reported in Table~\ref{tab:sigma2_linear1D}.
Again, the DF with each NN perform close to the KF. The DF with the CNN
does a little better while the one with the RNN falls behind.

\begin{table}[thpb]
\scalebox{0.95}{
\begin{tabular}{|l|l|l|l|l|l|l|}
\hline
$\sigma_0^{AM}$  & 0.1  & 0.5  & 1    & 1.5  & 2    & 2.5   \\ \hline
KF & 5.91  & 6.28 & 6.93 & 7.85 & 8.94 & 10.13 \\ \hline
DNN 1D Lin  & 6.48 & 6.57 & 7.09  & 8.25  & 9.88  & 11.74 \\ \hline
CNN 1D Lin                                                                                                              &  5.69 & 5.88 & 6.59  & 7.75  & 9.19  & 10.80 \\ \hline
RNN 1D Lin                                                                                                              &  6.11 & 6.32 & 7.32  & 9.33  & 12.12 & 15.38  \\ \hline
\end{tabular}
}
\caption{Relative Errors (in \%) for Different  $\sigma_0^{AM}$ (1D Linear Models)
  with $\sigma_0=\sigma_0^{NM}=2$}
\label{tab:sigma2_linear1D}
\end{table}

The results for the 2D linear model with
$ F^0_n = \big(\begin{smallmatrix}0 & 1\\1 & 1 \end{smallmatrix}\big)$ are
provided in Table~\ref{tab:sigma2_linear2D_1}.
The robustness of the DF is comparable to that of the KF.

\begin{table}[thpb]
\scalebox{0.95}{
\begin{tabular}{|l|l|l|l|l|l|l|}
\hline
$\sigma_0^{AM}$  & 0.1  & 0.5  & 1    & 1.5  & 2    & 2.5   \\ \hline
KF  & 6.87 & 6.97 & 7.26 & 7.73  & 8.35  & 9.07 \\ \hline
DNN 2D Lin & 5.61 & 5.82 & 6.45 & 7.44 & 8.68 & 10.07 \\ \hline
CNN 2D Lin & 8.49 & 7.55 & 7.09 & 7.49 & 8.48 & 9.81  \\ \hline
RNN 2D Lin & 6.60 & 6.63 & 6.98 & 7.76 & 8.89 & 10.26 \\ \hline
\end{tabular}
}
\caption{Relative Errors (in \%)
  for Different  $\sigma^{AM}_0$ (2D Linear Models) with
  $\sigma_0=\sigma_0^{NM}=2$}
\label{tab:sigma2_linear2D_1}
\end{table}

\para{Non-Linear Systems.}
We consider nonlinear models and compare the results of three kinds of neural networks with the corresponding extended Kalman filter.

\para{1D Non-Linear Model.}
First, we consider the following 1D model:
\begin{equation}\label{1dnlequ}
\left\{\begin{array}{l}
x_{n+1}=x_n+\eta \sin(5x_n)+\sqrt{\eta}\ \sigma u_n,\ x_0=1,\\
y_n= x_n+\sigma_0^{}v_n, \ n=0,1,2,\ldots,
\end{array}\right.\\
\end{equation}
with $u_n$ and $v_n$ which are independent Gaussian $N(0,1)$ random variables.

\para{2D Non-Linear Model.}
The 2D model is given by:
\begin{equation}\label{2dnlequ}
\left\{\begin{array}{l}
x_{n+1}=x_n+\eta \sin(5 F_n^0 x_n)+\sqrt{\eta}\  G^0_n u_n,\ x_0=[1,1]^T,\\
y_n= H'_n x_n+\sigma_0^{}v_n, \ n=0,1,2,\ldots,
\end{array}\right.\\
\end{equation}
with $u_n$ and $v_n$ which are independent 2D normal random variables.
We take $\eta=0.005$, $\sigma_0=0.5$, and
$ F^0_n = \big(\begin{smallmatrix}0 & 1\\1 & 1 \end{smallmatrix}\big)$.

  In Table \ref{tab:renl}, it can be seen that the DF with each NN outperforms
  the EKF in both the one-dimensional and two dimensional cases.
  
\begin{table}[thpb]
\centering
\begin{tabular}{|l|l|l|l|l|}
\hline
         & EKF  & DNN  & CNN  & RNN \\ \hline
1D Non-Linear  &10.04  & 6.16 & 5.68 &  6.40 \\ \hline
2D Non-Linear  &8.87  & 5.27 & 5.13 &  5.37    \\ \hline
\end{tabular}
\caption{Relative Errors (in \%) of Nominal Model for Non-Linear Models}
\label{tab:renl}
\end{table}


The corresponding robustness comparisons are reported in
Tables~\ref{tab:sigma2_nonlin1D} and \ref{tab:sigma2_nonlin2D_1}.
In both the one and two dimensional cases, the DF with each NN shows
similar robustness as the EKF except with the RNN which under performers the
rest.


\begin{table}[thpb]
\scalebox{0.9}{
\begin{tabular}{|l|l|l|l|l|l|l|}
\hline
$\sigma_0 ^{AM}$    & 0.1   & 0.5  & 1    & 1.5  & 2     & 2.5   \\ \hline
EKF  & 8.80 & 9.05 & 9.75 & 10.76  & 11.97 & 13.33  \\ \hline
DNN 1D NL  & 9.54 & 9.47 & 9.87  & 11.29 & 13.51 & 16.04  \\ \hline
CNN 1D NL                                                                                                              & 8.27 & 8.40 & 9.26  & 10.77 & 12.68 & 14.86 \\ \hline
RNN 1D NL                                                                                                              &  8.54 & 8.81 & 10.24 & 13.19 & 17.13 & 21.51  \\ \hline
\end{tabular}
}
\caption{Relative Errors (in \%) for Different  $\sigma_0^{AM}$
  (Non-Linear Models) with $\sigma_0=\sigma_0^{AM}=2$}
\label{tab:sigma2_nonlin1D}
\end{table}

\begin{table}[thpb]
\scalebox{0.9}{
\begin{tabular}{|l|l|l|l|l|l|l|}
\hline
$\sigma_0 ^{AM}$    & 0.1   & 0.5  & 1    & 1.5  & 2     & 2.5   \\ \hline
EKF  & 9.19 & 9.34 & 9.78  & 10.44 & 11.28 & 12.25  \\ \hline
DNN 2D NL & 7.08  & 7.31 & 8.07 & 9.32 & 10.95 & 12.81 \\ \hline
CNN 2D NL & 11.12 & 9.84 & 9.12 & 9.53 & 10.74 & 12.40 \\ \hline
RNN 2D NL &  7.82 & 7.88 & 8.33 & 9.34  & 10.74 &12.43 \\ \hline
\end{tabular}
}
\caption{Relative Errors (in \%) for Different  $\sigma_0^{AM}$
  (Non-Linear Models) with $\sigma_0=\sigma_0^{NM}=2$}
\label{tab:sigma2_nonlin2D_1}
\end{table}

\begin{figure}[thpb]
	\begin{center}
		\includegraphics[width=1\linewidth]{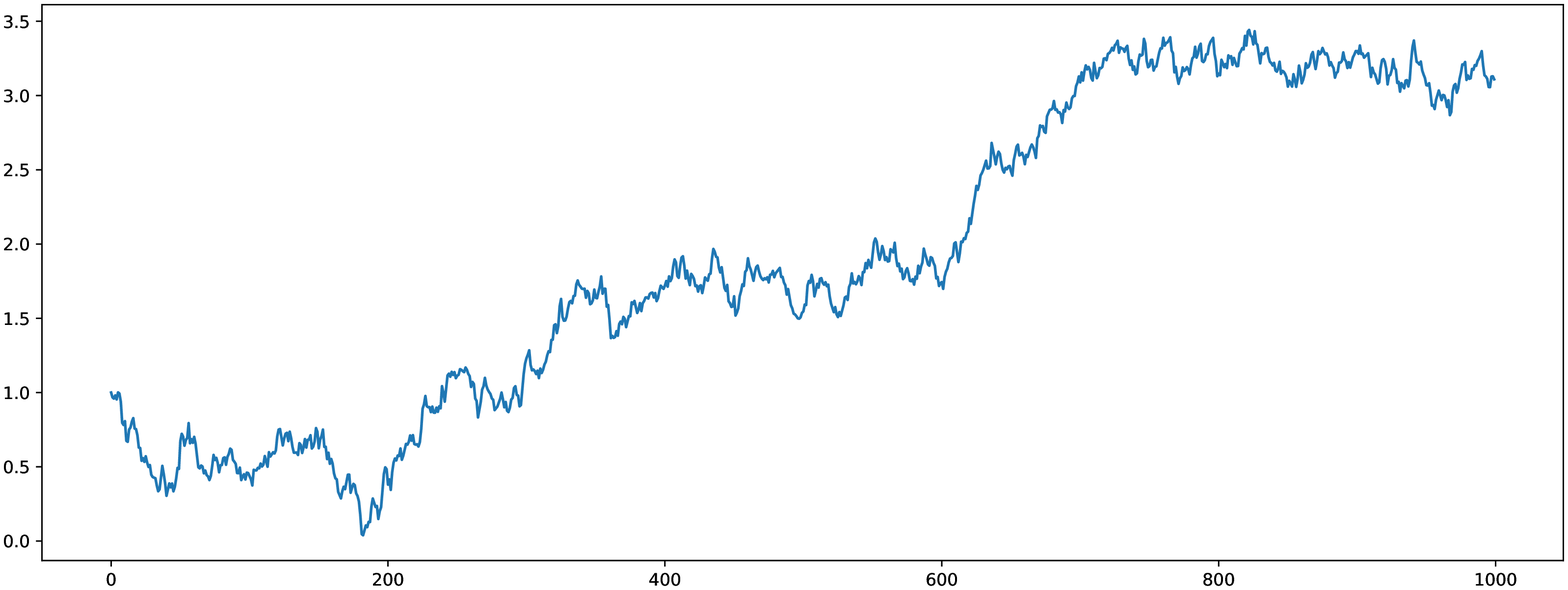}
		\caption{A Sample Path of $x_n$ in 1D Non-Linear Model}
		\label{fig:path1890}
	\end{center}
\end{figure}

\begin{figure}[thpb]
	\begin{center}
		\includegraphics[width=1\linewidth]{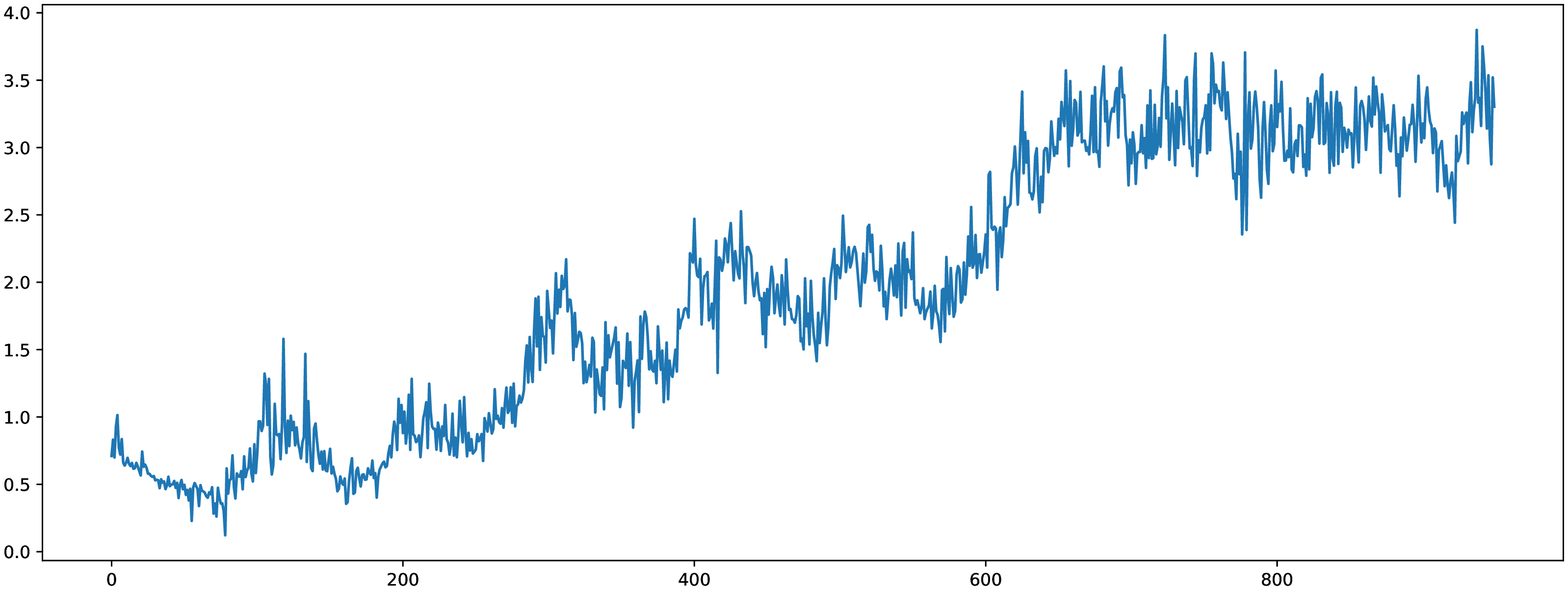}
		\caption{The Corresponding DF $\widetilde x_n$ with DNN}
		\label{fig:dnnpred}
	\end{center}
\end{figure}

\begin{figure}[thpb]
	\begin{center}
		\includegraphics[width=1\linewidth]{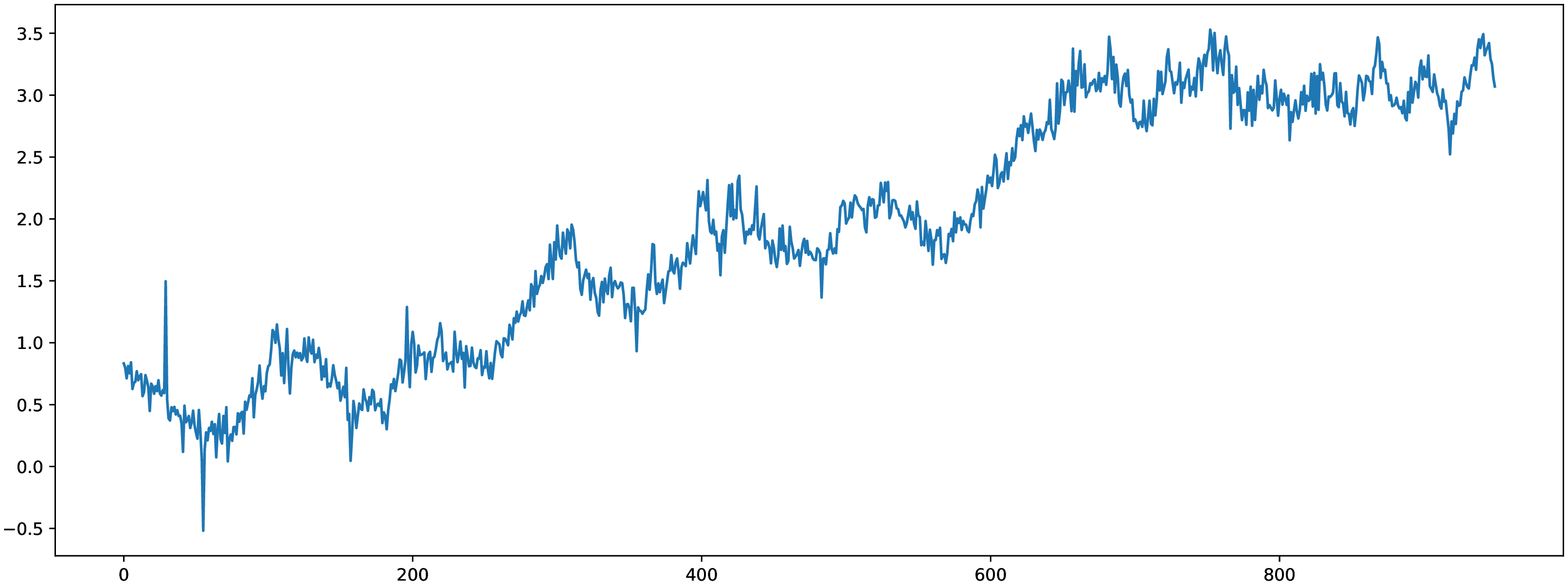}
		\caption{The Corresponding DF $\widetilde x_n$ with CNN}
		\label{fig:cnnpred}
	\end{center}
\end{figure}

\begin{figure}[thpb]
	\begin{center}
		\includegraphics[width=1\linewidth]{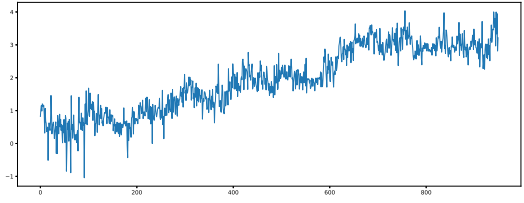}
		\caption{The Corresponding DF $\widetilde x_n$ with RNN}
		\label{fig:rnnpred}
	\end{center}
\end{figure}

In Figures~\ref{fig:path1890}-\ref{fig:rnnpred},
a sample of
$x_n$ in the one-dimensional system with $\sigma_0 =\sigma_0^{NM}=2$ is given along with the corresponding
DF $\widetilde x_n$ with DNN, CNN, and RNN. According to these pictures,
it appears that the DF with CNN provides smoother path. This is
consistent with the results included in Table~\ref{tab:renl}.

\para{Switching Models.}
Next, we consider a switching model with jumps and apply the DF
to these models. For the jump case,
neither the KF nor the EKF can be applied
due to the presence of the switching process and lack of recursive
dynamic equations.

\para{1D Switching Model.}
As in Wang et al.\cite{WangYZ}
we consider the following model:
\beq {1dequ}
\left\{ \begin{array}{l}
x_n=\sin(n\eta\al_n+\sigma u_n),\ n=0,1,2,\ldots,\\
y_n=x_n+\sigma_0 v_n.
\end{array}\right.
\eeq

We take $\al(t)\in\{1,2\}$ to be a continuous-time Markov chain
with generator 
$Q=\big(\begin{smallmatrix} -2 & 2 \\ 2 & -2 \end{smallmatrix}\big)$,
Using step size $\eta=0.005$ to discretize $\al(t)$ to get
$\al_n=\al(n\eta)$.
We also take $\sigma=0.1$ and $\sigma_0=0.3$.

\para{2D Switching Model.}
We consider the following 2D switching system:
\[
\left\{\begin{array}{l}
x_n=\sin(n\eta\al_n+\sigma u_n),\ n=0,1,2,\ldots,\\
\bar x_n=\cos(n\eta\al_n+\sigma u_n),\ n=0,1,2,\ldots,
\end{array}\right.
\]
The observation 
\begin{equation}\label{swtequ}
    \begin{pmatrix}y_n \\ \bar y_n \end{pmatrix} 
    = H \begin{pmatrix} x_n \\ \bar x_n \end{pmatrix} + \sigma _0 u_n,
\end{equation}
with $u_n$ as 2D white noise. We take $\eta=0.005$, $\sigma=0.3$.
and $ H = \big(\begin{smallmatrix}0 & 1\\1 & 1 \end{smallmatrix}\big)$.
We choose nominal model $\sigma^{NM}_0 = 2$ for the NNs training,

  In these examples, the KF (or the ETF) would not apply due to lack 
  of recursive dynamic systems. We only compare the robustness of the DF
  with various NNs.  
  In Table~\ref{tab:sigma2_swt_1D}, the DF with the DNN exhibits better
  robustness while the one with the RNN underperforms.

\begin{table}[thpb]
\scalebox{0.9}{
\begin{tabular}{|l|l|l|l|l|l|l|}
\hline
$\sigma_0 ^{AM}$                                                                                                              & 0.1   & 0.5   & 1     & 1.5   & 2     & 2.5   \\ \hline
DNN 1D Swt                                                                                                              & 12.67 & 12.64 & 12.92 & 14.28 & 16.79 & 19.62 \\ \hline
CNN 1D Swt                                                                                                              & 18.01 & 17.34 & 16.90 & 17.74 & 19.80 & 22.52 \\ \hline
RNN 1D Swt                                                                                                              & 13.81 & 14.23 & 16.45 & 21.31 & 27.61 & 33.92 \\ \hline
\end{tabular}
}
\caption{Relative Errors (in \%) for Different  $\sigma_0^{AM}$
  (1D Switching Model) with $\sigma_0=\sigma_0^{NM}=2$}
\label{tab:sigma2_swt_1D}
\end{table}

The robustness of the DF under the two-dimensional switching model
is given in Table~\ref{tab:sigma2_swt_2Da}.
In this case, the DNN outperforms while the RNN drags much behind.

\begin{table}[thpb]
\scalebox{0.9}{
\begin{tabular}{|l|l|l|l|l|l|l|}
\hline
$\sigma_0 ^{AM}$                                                                                                              & 0.1   & 0.5   & 1     & 1.5   & 2     & 2.5   \\ \hline
DNN 2D Swt& 37.24 & 36.99 & 36.92 & 37.83 & 39.29 & 40.75 \\ \hline
CNN 2D Swt  & 16.36 & 15.38 & 15.48 & 17.16 & 19.68 & 22.47 \\ \hline
RNN 2D Swt  & 12.01 &  11.94  & 11.85  & 12.66 &  17.27  & 99.37      \\ \hline
\end{tabular}
}
\caption{Relative Errors (in \%) for Different  $\sigma_0^{AM}$
  (2D Switching Models) with $\sigma_0=\sigma_0^{NM}=2$}
\label{tab:sigma2_swt_2Da}
\end{table}

\begin{figure}[thpb]
	\begin{center}
		\includegraphics[width=0.98\linewidth]{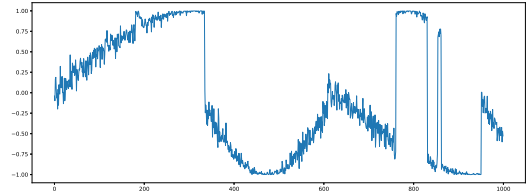}
		\caption{A Sample Path of $x_n$ in the 1D Switching Model }
		\label{fig:swt1980}
	\end{center}
\end{figure}

\begin{figure}[thpb]
	\begin{center}
		\includegraphics[width=0.98\linewidth]{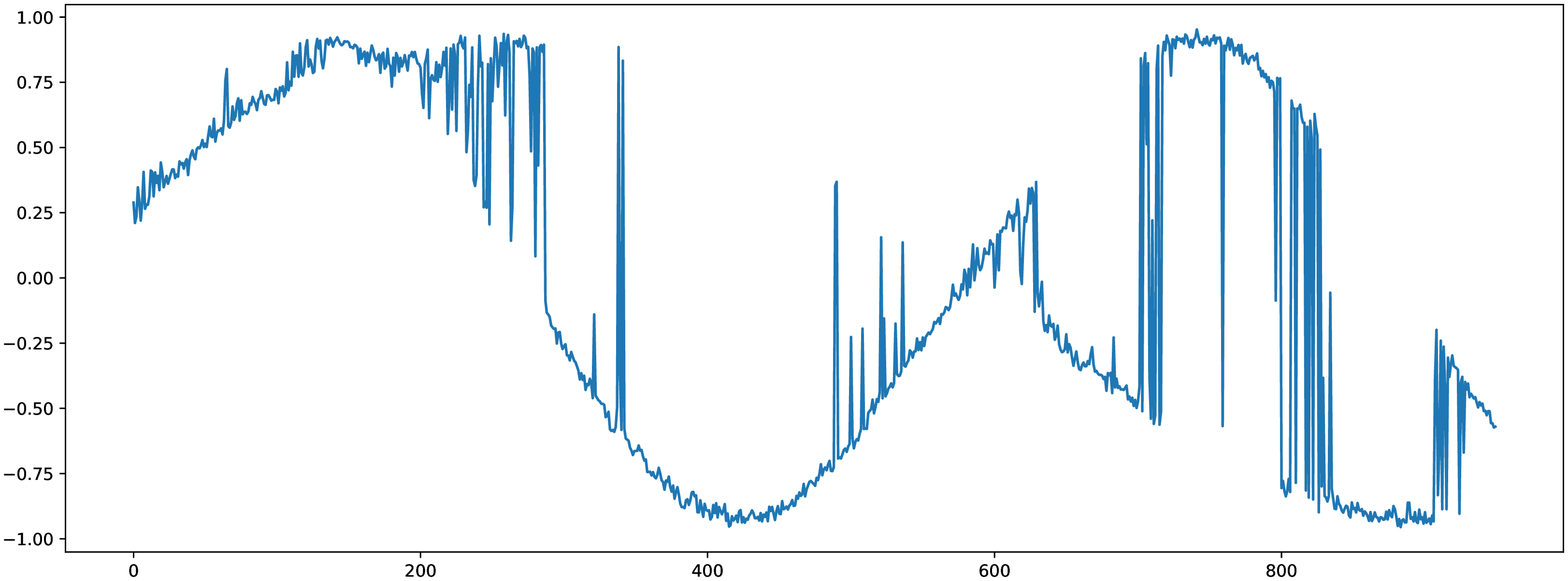}
		\caption{The Corresponding DF $\widetilde x_n$ with DNN}
		\label{fig:swtdnn}
	\end{center}
\end{figure}

\begin{figure}[thpb]
	\begin{center}
		\includegraphics[width=0.98\linewidth]{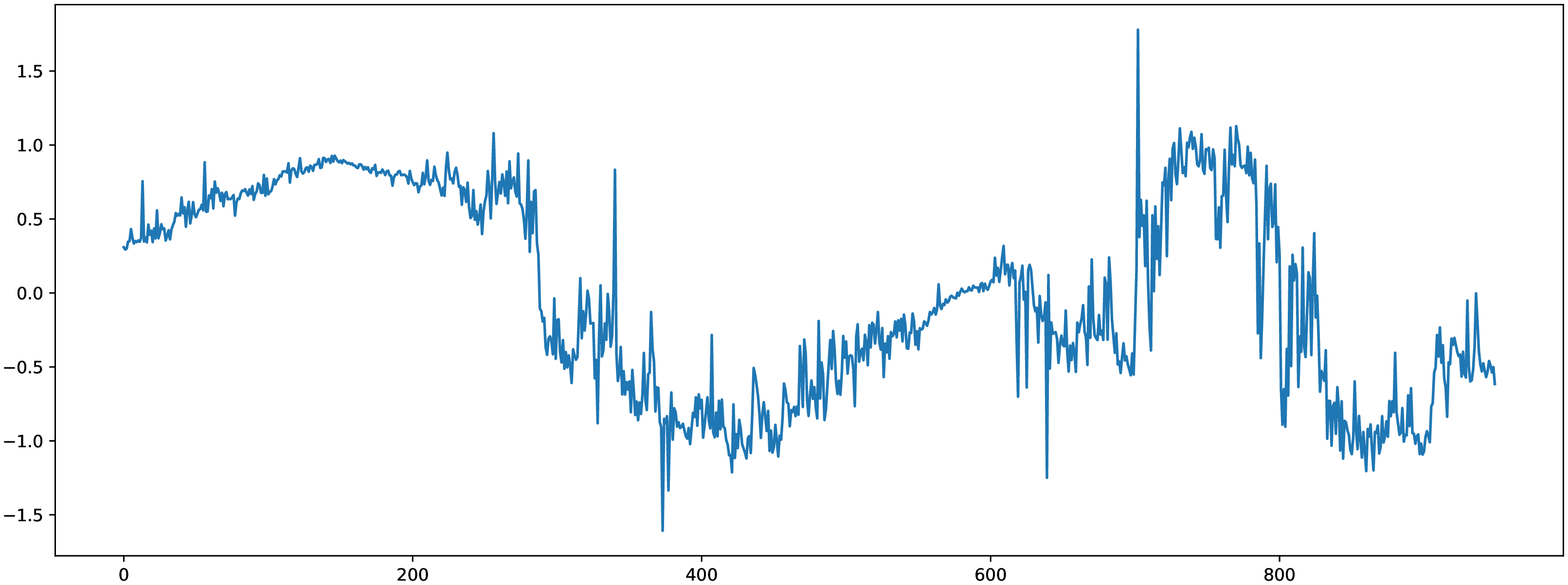}
		\caption{The Corresponding DF $\widetilde x_n$ with CNN}
		\label{fig:swtcnn}
	\end{center}
\end{figure}

\begin{figure}[thpb]
	\begin{center}
		\includegraphics[width=0.98\linewidth]{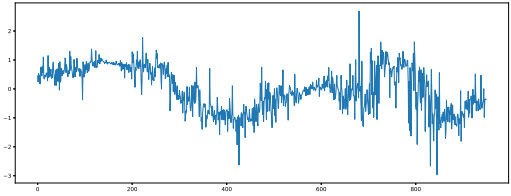}
		\caption{The Corresponding DF $\widetilde x_n$ with RNN}
		\label{fig:swtrnn}
	\end{center}
\end{figure}

In Figures~\ref{fig:swt1980}-\ref{fig:swtrnn},
  a sample of
  $x_n$ in the one-dimensional switching model with
  $\sigma_0^{NM}=\sigma_0^{AM}=2$
  is plotted
  along with the corresponding
  DF $\widetilde x_n$ with DNN, CNN, and RNN.
  In this case, the CNN appears doing better than others.

\begin{remark}
{\rm
  Note that the RNN performs poorly when $\sigma_0^{AM}=2.5$.
  This appears due to the LSTM's over fitting feature in the 2D case.
  One way to reduce such over fitting is to stop early to avoid loss function
  blowups.
  By and large, the early stopping will halt the NN's training by monitoring
  the loss function and exit when the loss value no longer improves
  in several consecutive epochs. We refer the reader to Jason \cite{BJason} for
  further details. Using such an early stopping technique, the corresponding
  relative errors are given in Table~\ref{tab:estopping_swt}.

\begin{table}[thpb]
\scalebox{0.9}{
\begin{tabular}{|l|l|l|l|l|l|l|}
\hline
$\sigma_0 ^{AM}$                                                                                                             & 0.1   & 0.5  & 1    & 1.5   & 2     & 2.5   \\ \hline

RNN 2D Swt & 12.06  & 11.89 & 11.81 & 12.56 & 14.54 & 17.73 \\ \hline

\end{tabular}
}
\caption{Relative Errors (in \%) of Early Stopping for Different $\sigma_0^{AM}$ (Switching Models) with $\sigma_0 =\sigma_0^{NM}=2$}
\label{tab:estopping_swt}
\end{table}

  }
  \end{remark}

\begin{remark}
{\rm
  In this paper, an iMac desktop with Intel Core i5 processor (3.2 GHz Quad-core, 24G DDR3) with Keras library of Python 3 was
  used for all numerical experiments.
In Table~\ref{tab:time},      the CPU time for 
one step calculation with various network architectures are given.
The DNN and CNN consume similar amount of time, while the RNN runs
much longer.
\begin{table}[thpb]
\centering
\scalebox{0.85}{
\begin{tabular}{|l|l|l|l|l|l|l|}
\hline
Time & 1D Lin & 1D NL & 1D Swt & 2D Lin & 2D NL & 2D Swt \\ \hline
DNN  &   11.0    &  10.5 &  11.1  &   14.1   &   13.5  &  14.7  \\ \hline
CNN  &  12.5    &  12.1 &   12.5 &   19.5    &   18.6  &  18.8  \\ \hline
RNN  &  17.5   &   17.5 &    10.6 &  185.5    &   175.0 &  182.2  \\ \hline
\end{tabular}
}
\caption{Running Time (in seconds)}
\label{tab:time}
\end{table}

We would like to comment that the network training is the most time
consuming part. Nevertheless, this will not affect much in real time
filtering because the training can be done offline.
    }
  \end{remark}
  
\section{Conclusion}\label{sec:conclusion}

In this paper, we have studied the deep filter with different neural network
architectures. The DF performs  similarly as the Kalman filter in linear
models and outperforms the KF in nonlinear models,
In addition, the CNN outperforms the DNN and RNN on average.
The DF is applicable to general dynamic systems and it does not
require any model specifications. 
To implement the DF, real data can be used directly to
train the deep neutral network. Therefore, model calibration can be
eliminated all together.

\section*{Acknowledgments}
This research was supported in part by the Simons Foundation (849576).


%





\end{document}